\documentclass[aps,twocolumn,showpacs]{revtex4}
\usepackage{dcolumn}
\usepackage{graphicx}
\usepackage{amsmath}
\usepackage{amsfonts}
\usepackage{amssymb}
\usepackage{psfrag}
\usepackage{wrapfig}
\usepackage{subfigure}
\usepackage{makeidx}
\usepackage{bm}
\usepackage{epsf}

\begin{document}

\title{Tunneling Dynamics Between Atomic Bright Solitons}

\author{Li-Chen Zhao$^{1,2}$}
\author{Liming Ling$^{3}$}
\author{Zhan-Ying Yang$^{1,2}$}
\author{Wen-Li Yang$^{2,4}$}


\address{$^1$School of Physics, Northwest University, Xi'an 710069, China}
\address{$^{2}$Shaanxi Key Laboratory for Theoretical Physics Frontiers, 710069, Xi'an, China}
\address{$^{3}$School of Mathematics, South China University of Technology, 510640, Guangzhou, China}
\address{$^{4}$Institute of Modern Physics, Northwest University, 710069, Xi’an, China}

\begin{abstract}
We investigate tunneling behavior between two bright solitons in a Bose-Einstein condensate with attractive contact interactions between atoms.  The explicit tunneling properties including tunneling particles and oscillation period are described analytically, which indicates that the periodic tunneling form is a nonlinear Josephson type oscillation. The results suggest that the breathing behavior of solitons comes from the tunneling mechanism in an effective double-well potential, which is quite different from the modulational instability mechanism for Akhmediev breather and K-M breather. Furthermore, we obtain a phase diagram for two soliton interaction which admits tunneling property, particle-like property, interference property, and a resonant interaction case. The explicit conditions for them are clarified based on the defined critical distance $d_c$ and spatial interference period $D$.

\end{abstract}
\maketitle

\section{Introduction}
\label{intro}
 Soliton has been a well known nonlinear localized wave for its particle-like property, which can exist in many different physical systems \cite{Zabusky,Segev,Serkin, Serkin1,Zhao,Akhmediev2}.  However, it is
fundamentally a wave packet, which should admit wave properties, such as tunneling behavior and interference behavior. They can
interfere with each other during their interaction process \cite{Snyder,Kumar,Helm}. Recently, bright matter wave solitons' interference fringe was demonstrated experimentally \cite{McDonald}. Soliton-based matter-wave interferometer was proposed theoretically in a harmonic potential trap with a Rosen-Morse barrier at its
center \cite{Polo}, or with a local nonlinear repulsive potential \cite{Boris}. We have obtained the properties of interference pattern analytically and exactly  between bright solitons with nonzero relative velocity in \cite{Zhao3}. The results suggested that the interference properties can be used to measure soliton's velocity and nonlinear coefficient in nonlinear Schr\"{o}dinger equation (NLSE) described systems.  The apparent repulsion between solitons with
relative phase $\pi$ is actually an interference pattern generated by the two solitons passing
through each other, which was verified in a recent experiment \cite{Nguyen}. It is believed that soliton interaction has a bright future in
precision-measurement experiments \cite{Billam3}. On the other hand,  the tunneling dynamics of soliton have been discussed mainly for the cases that one solion collides on external potential barriers or interfaces \cite{T1,T2,Martin,Tkesh}.  Moreover, the tunneling dynamics can also happen between two bright solitons, since two bright solitons induce a time-dependent double well through nonlinear interactions \cite{Karamatskos}. The tunneling behavior brings the atom exchanges between bright solitons, which has been observed numerically  in \cite{Salasnich}. However, the tunneling dynamics between bright solitons have not been studied systematically, as far as we know. Here we intend to study on this nonlinear tunneling dynamics in the self-induced double-well potential, since the tunneling dynamics between bright solitons should be distinctive from those cases for which one solion collides on external potential barriers or interfaces  \cite{T1,T2,Martin,Tkesh}.

In this paper, we study the tunneling behavior between two bright solitons in a Bose-Einstein condensate(BEC) system, since BEC is a macroscopic quantum state and it is convenient to realize bright soliton \cite{BEC,Billam}. We demonstrate that the breathing behavior of bound state solitons can be understood well by the tunneling dynamics of atoms in an effective double-well potential induced by the two solitons.  This can be seen as the mechanism for breathing bound state of two solitons, in contrast to the modulational instability mechanism for Akhmediev breather and K-M breather. The tunneling rate and oscillation period are derived analytically, which suggest that the periodic tunneling form is a nonlinear Josephson type oscillation. The results provide new possibilities to observe nonlinear Josephson tunneling of atoms based on two bright matter wave solitons with the same initial velocity. Furthermore, we obtain a phase diagram for soliton interaction according to the relative velocity and distance between solitons, which mainly including four distinctive cases: visible tunneling, no visible tunneling and interference,  visible interference, and a resonant interaction case. The explicit conditions for them are clarified based on the defined critical distance $d_c$ and spatial interference period $D$.

Our presentation of the above features will be structured
as follows. In Sec. II,  we investigate tunneling behavior between two atomic bright solitons with identical velocities, based on the well-known two-soliton solution of the simplest NLSE.  We explain the breathing behavior through a tunneling dynamics analysis in an effective double-well potential seen from quantum mechanics viewpoint. Furthermore, we give some analytical expressions to describe the tunneling dynamics quantitatively, and define a critical distance $d_c$ between solitons under which the tunneling behavior is visible.  In
Sec. III , the tunneling behavior between solitons in all other possible cases are studied in detail. We obtain a phase diagram for soliton interactions, which admits four distinctive cases. The explicit conditions for them are clarified based on the critical distance $d_c$ between solitons and spatial interference period $D$.  Finally,  we summarize the results and
present our conclusions in Sec. IV.

\section{The tunneling behavior of two bright solitons with Identical velocities}
\label{sec:1}
It has great significance in practice to study soliton dynamics in a BEC system, since BEC is a macroscopic quantum state and the well-developed density and phase modulation techniques make it be convenient to excite bright soliton with arbitrary profile and phase \cite{McDonald,Becker}. We consider a cigar-shaped BEC with a weak harmonic trapping potential $\gamma \ x^2$ ($\gamma =\omega_0/(g\omega_{\perp})$) along the cigar direction $x$. $\omega_{\perp}$ and $\omega_0$ are corresponding harmonic oscillator frequencies, $m$ is the atom mass and the Feshbach nonlinear coefficient is $g=|a_s|/a_B$ ($a_B$ is the Bohr radius) and $a_s$ is the scattering length between atoms \cite{BEC}. The bright soliton in BEC has been created with parameters $N\approx \times 10^3  $, $\omega_{\perp}=2\pi \times700 Hz $, $\omega_0=2 \pi \times 7 Hz$, and $a_{s}=-4 a_B$ for $^7Li$ \cite{BsE}.
With $\gamma =\omega_0/(g\omega_{\perp}) \ll 1$, it is proper to discuss the dynamics of solitons around the place $x=0$ with ignoring the effects of trapping potential. Then the dynamics of condensate wave function can be described by the following NLSE with scalar units
\begin{equation}
i\frac{\partial U(x,t)}{\partial t}= -\frac{\partial^2
}{\partial x^2}\ U(x,t)-2 |U(x,t)|^2 \ U(x,t).
\end{equation}
Time $t$ and coordinate $x$ are measured in units of $2g /\omega_{\perp}$ and $\sqrt{g} a_{\perp}$, where $a_{\perp}=(\hbar/ m \omega_{\perp})^{1/2}$ is linear oscillator length in the transverse directions, respectively. The system admits bright soliton and the interaction between solitons can be studied based on exact multi-soliton solution which can be obtained by B\'{a}cklund transformation \cite{Mat,Ling}. It has been well known that two bright solitons with zero relative velocity can form a breathing bound state \cite{Segev}. Solitons with controlled relative phase have been realized in experiments \cite{Billam}. Here, we firstly investigate the breathing behavior between solitons based on the two-soliton solution with zero initial  relative velocity.
\begin{figure}[htb]
\centering
\includegraphics[height=70mm,width=90mm]{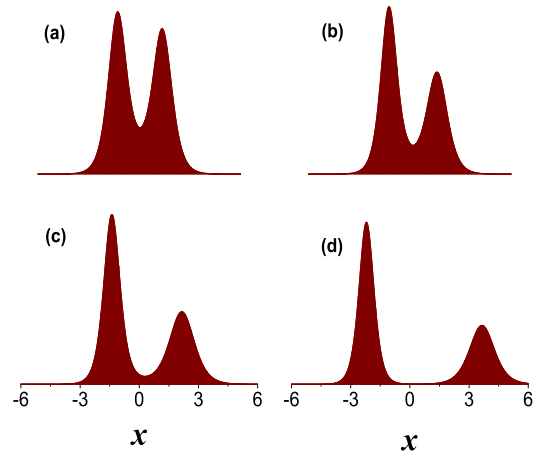}
\caption{(color online) The initial density profiles of the solitons with different distance between solitons. The overlapping of them become less and less for the ones in (a-d) with larger initial soliton separations. We will observe the tunneling dynamics between them in Fig 2.  (a) $-c_1=c_2=0.05$, (b) $-c_1=c_2=0.25$, (c) $-c_1=c_2=0.6$, (d) $-c_1=c_2=1.5$. The other parameters are $a_1 = 0.6$, $a_2 = 1$, $d_1 = d_2 = 0$.} \label{fig2}
\end{figure}

\begin{figure}[htb]
\centering
\includegraphics[height=70mm,width=85mm]{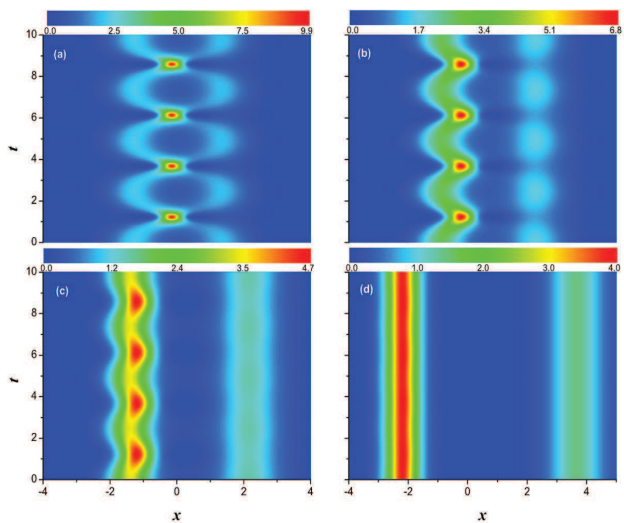}
\caption{(color online) The evolution of two bright solitons with different overlapping degrees from the initial conditions in Fig. 1.  It is seen that there are both location and peak oscillations for solitons with strong overlapping. The oscillation behavior become weak with weak overlapping. The peak oscillations of two solitons indicate that there are particle exchanges between two solitons. } \label{fig1}
\end{figure}

We consider the tunneling behavior of solitons with zero initial relative velocity and different distance between them. The initial profiles for them are shown in Fig. 1(a-d). The evolutions of these different initial conditions are shown in Fig. 2(a-d) respectively.
The dynamic of two solitons can be described by the well-known two-soliton solution
\cite{Mat,Ling}, $U(x,t)=\frac{{\rm 4 i} \left( {a_{{1}}}^{2}-{a_{{2}}}^{2} \right)  F_1(x,t)}{F_2(x,t)}$,
where $F_1(x,t)= a_{{1}}\cosh
 \left( 2 a_{{2}}x+2 c_{{2}} \right) {{\rm e}^{4{\rm i} {a_{{1}}}^{2}t+ i d_1}}
-a_{{2}}\cosh \left( 2 a_{{1}}x+2 c_{{1}} \right) {{\rm e}^{4{\rm i} {a_{{
2}}}^{2}t+i d_2}} $, and $F_2(x,t)=\left( a_{{1}}+a_{{2}} \right) ^{2}\cosh[2( a_{{1}}-a_
{{2}})x+2(c_{{1}}-c_{{2}})] + \left( a_{{1}}-a_{{2}}
 \right) ^{2}\cosh[2\left( a_{{1}}+a_{{2}} \right) x+2(c_{{
1}}+c_{{2}})] -4 a_{{1}}a_{{2}}\cos \left( 4  \left( {a_{{1
}}}^{2}-{a_{{2}}}^{2} \right) t + d_1-d_2\right)$.
The parameters $a_1$ and $a_2$ determine peak value of solitons respectively,  $c_1$ and $c_2$ determine the initial locations of solitons. $d_1$ and $d_2$ can be used to vary the relative phase between solitons. It is pointed that the soliton solution can be shifted on time or space without affecting the essential dynamics of them through coordinate shift operation. In this paper, we perform related operations to demonstrate the dynamics conveniently on temporal direction. We firstly discuss on the case with $a_1\neq a_2 $, the case with $a_1=a_2$ corresponds to the resonant interaction between solitons and it will be discussed in the Sec. III (B) part.

It is seen that there are both location and peak oscillations for solitons with strong overlapping. The oscillation behavior
become weak with weak overlapping.  The overlapping part of solitons become less and less with increasing the distance of soliton and keeping the other parameters determining the profile of each soliton unchanged. Combining the dynamics results in Fig. 2, we can see that breathing behavior become weaker with less overlapping part.  This suggests that the overlapping part plays essential role in the breathing behavior of the two solitons. Two solitons with zero initial relative velocity and small separations can form a bound state which breath with time evolution periodically (see Fig. 2(a)). Similar breathing bound states formed by solitons have been shown widely \cite{Segev}.  Moreover, if the two solitons are initially located with larger distance, the location oscillation will become less obvious, and the soliton's peak value will demonstrate oscillation (see Fig. 2 (b) and (c)). This suggests that there are particles exchange between the two bright solitons. If we further increase the distance between solitons, the breathing behavior including both location and peak breathing dynamics will become invisible, as shown in Fig. 2 (d). Nextly, we will try to understand the breathing behavior between solitons from a physical mechanism.

The breathing peak of solitons suggest there are atoms exchange between solitons. This makes us think about the tunneling dynamics of matter wave in a double well potential in quantum theory \cite{Joseph}. In fact, the nonlinear term $-2 |U(x,t)|^2$ in Eq. (1) can be seen as a potential $V(x,t)$ in quantum dynamics \cite{Karamatskos}. In quantum dynamics, the wave function will denote the spatial probability distribution function of one-particle. The norm of wave function will denote probability distribution. This naturally corresponds to the density distribution for large number of identical particles occupying on the same quantum state. For the BEC system, this is reasonably the density distribution of atoms. The oscillation behavior has two parts: peak oscillation and  location oscillation.

\begin{figure}[htb]
\centering
\includegraphics[height=70mm,width=90mm]{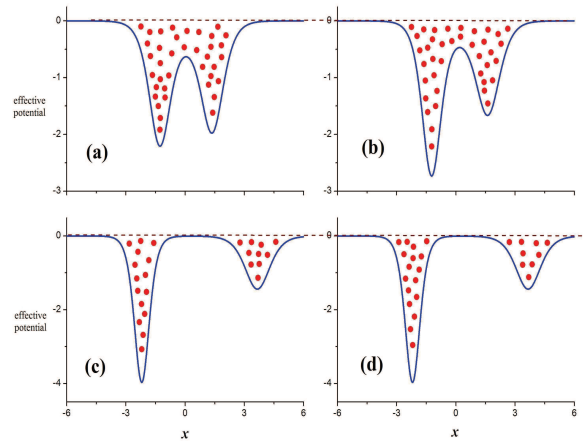}
\caption{(color online) The effective double-well potential of the initial bright soliton condition, which are seen from the quantum mechanics viewpoint. It is natural to expect that there should be tunneling behavior between the effective two wells, which can be used to understand the breathing behavior of solitons. The less overlapping of solitons makes the effective barrier between two double well higher and wider, which restrains the tunneling behavior.} \label{fig3}
\end{figure}

\subsection{The soliton amplitude breathing behavior}
Firstly, we investigate the peak oscillation (as the case shown in Fig. 2 (c)), for which solitons show obvious breathing peak and the locations are nearly fixed. The corresponding potential well of the initial solitons in Fig. 2(c) is shown in Fig. 3(c). We can see that the two bright solitons induce an effective double-well structure potential. Then, it is natural to expect that there should be some periodic tunneling behavior between the two wells, based on quantum tunneling theory \cite{Joseph}.  Fig. 2(c) indeed shows that the left soliton demonstrates a tunneling behavior to the other. The occupying probability in each well for a single-particle will corresponds to the particles number in each well for atoms in BEC.
Based on the tunneling mechanism, we can know that the tunneling behavior becomes weaker and weaker with overlapping of solitons is less and less. This comes from that the less overlapping of solitons makes the effective barrier between two double well be higher and wider (see Fig. 1(d) and Fig. 3(d)), which restrains tunneling behavior. This can be used to explain that the tunneling behavior of the ones in Fig. 2(d) is almost invisible. When the overlapping part is large, the corresponding effective barrier between two wells will become lower and narrower (Fig. 3(a) and (b)). This brings the tunneling behavior become more drastic (see Fig. 2(a) and (b)).  Therefore, the breathing behavior of two parallel solitons comes from the tunneling mechanism.

Two bright solitons with zero relative velocity can interact with each other and form a breathing bound state \cite{Segev}. The breathing behavior is similar to the K-M breather \cite{K-M}. However, the mechanism for this breathing behavior should be different from the K-M breather \cite{K-M}, since they exist on different backgrounds. K-M breather comes form the modulational instability, which mainly involves the perturbation signal interact with a plane wave background. The breathing bound state here is formed by two bright soliton on a zero background. One can prove that the bright soliton does not admit modulational instability.  It should be noted that two bright solitons can induce an effective double-well potential for a NLSE described system. We demonstrate that the tunneling mechanism can be used to explain the breathing behavior between solitons, which is different from the modulational instability mechanism for AB and K-M breather.

\subsection{The soliton location breathing behavior}
Then, let us discuss the location breathing behavior of solitons.
It should be noted that the double-well potential is self-induced by the distribution of atoms ($-2 |U(x,t)|^2$ where $U(x,t)$ is the two-soliton solution given above). The structure of double-well evolves simultaneously with the evolution of bright solitons, since atoms tunneling from one soliton to the other change the double well structure synchronously. Therefore, we call it as the tunneling behavior of matter wave in a self-induced double-well potential, in contrast to the external double-well potential in usual quantum theory. This nonlinear interaction effect brings the soliton's location oscillate with time.  Meanwhile, the overlapping of them is heavier, which makes the effective barrier between double well become much lower and narrower (see Fig. 3(a) and (b)). This makes the tunneling behavior happen greatly. Therefore, the ones in Fig. 2(a) and (b) demonstrate obvious location oscillation and peak oscillation simultaneously. Bright solitons in Fig. 2(c) and (d) have much larger distance and weak overlapping between them, which makes the nonlinear interaction between the two solitons much weak.  This brings the ones in Fig. 2(c) demonstrate weak location oscillation, and the ones in Fig. 2(d) admit nearly invisible location oscillation.

Moreover, the quantum tunneling and wave-particle duality of soliton have been discussed well in \cite{wavep,Serkin4,T1,T2,WL}. Most of them are discussed by studying reflection and transmission coefficients of a soliton colliding on a barrier potential or an interface \cite{Tkesh}. It has been shown that there are some exotic dynamics for soliton tunneling behavior. We emphasize that tunneling dynamics here is  between bright solitons with no external potential barrier or interface, which is distinctive from the previous studies \cite{wavep,Serkin4,T1,T2,WL,Tkesh}. The effective double-well potential is induced by two bright solitons, and changes with evolution of tunneling dynamics.

\begin{figure*}[htb]
\centering
\includegraphics[height=65mm,width=165mm]{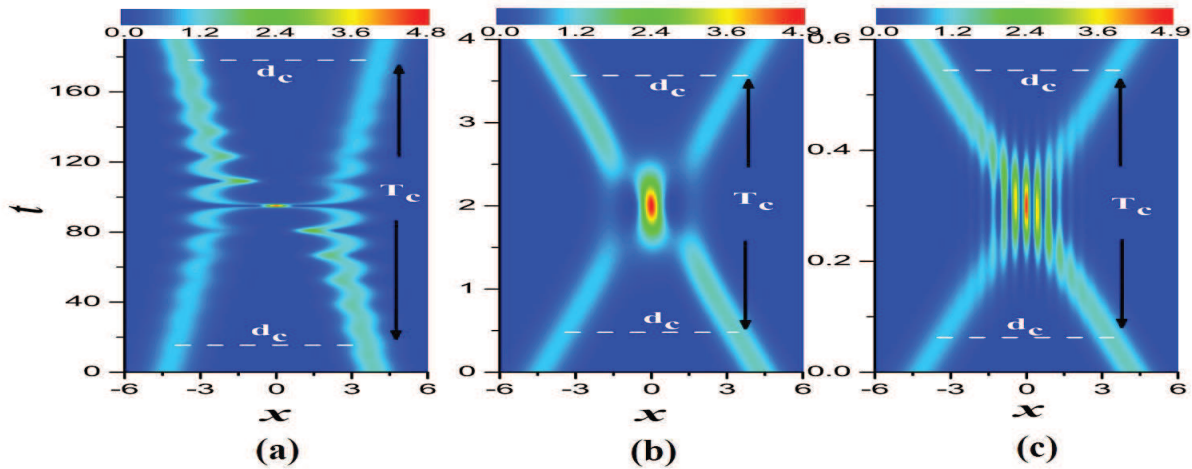}
\caption{(color online) The interaction between two solitons with different group velocities.  (a) for the case with visible tunneling behavior, for which the relative velocity of solitons is small ($-b_1 =b_2= 0.005$).  (b)for the case with no visible tunneling behavior and no visible interference pattern, for which the relative velocity of is not too small or large ($-b_1 =b_2= 0.5$). (c) for the case with visible interference pattern, for which the relative velocity of solitons is large ($-b_1 =b_2= 3.5$).
How small or large is proper for the tunneling behavior or interference pattern is given explicit criterions to make judgment.  The other parameters are $a_1 = 0.6$,  $a_2 = 0.5$, $c_1 = c_2 = 0$, $d_1 =d_2 =0$.} \label{fig3}
\end{figure*}

\subsection{An analysis on tunneling behavior and the defined critical distance}
It is shown that the tunneling behavior becomes weaker with the distance between two solitons is larger.  How far is the distance between the two solitons  proper for the visible tunneling behavior?  We would like to discuss on this through defining a criterion for the visible tunneling behavior. Since we intend to find how far the largest distance is proper for visible tunneling behavior, the far distance can be used to analyze the two-soliton solution asymptotically. Assuming that $a_1>a_2>0$, $c_1>0$, $c_2<0$ and $|c_1-c_2|$ is enough large which means that the distance between solitons is far, we can obtain the asymptotical analysis under the defined conditions. The left soliton is along the line
\begin{equation*}
  x_{left}=\frac{1}{a_{1}}\left[\frac{1}{2}\ln  \left( {\frac {a_{1}-a_{2}}{a_{1}+a_
{2}}} \right)-c_{1}\right],
\end{equation*}
and the right one is along the line
\begin{equation*}
 x_{right}=-\frac{1}{a_{2}}\left[\frac{1}{2}\ln  \left( {\frac {a_{1}-a_{2}}{a_{1}+a_
{2}}} \right)+c_{2}\right].
\end{equation*}
The distance between two solitons can be calculated as $d=|x_{right}-x_{left}|$, which can be used to evaluate the proper distance for visible tunneling behavior.   $|q^2|(x_{left},t)\approx 4a_1^2=P_1$ and $|q^2|(x_{right},t)\approx 4a_2^2=P_2$ denote the peak values of the two solitons respectively.
It is essential to calculate the particle numbers of two solitons to characterize the tunneling behavior explicitly.  To calculate the particle number of each soliton conveniently, we set
\begin{equation*}
  c_2=-\frac{1}{2a_{{1}}}\left[ 2\,a_{{2}}c_{{1}}-a_{{2}}\ln  \left( {\frac {a_{{1}}-a_{
{2}}}{a_{{1}}+a_{{2}}}} \right) +a_{{1}}\ln  \left( {\frac {a_{{1}}-a_
{{2}}}{a_{{1}}+a_{{2}}}} \right)  \right].
\end{equation*}
This condition can ensure that $x=0$ is the central point between the two solitons. And the distance between solitons' center in this setting can be calculated as $d=|x_{right}-x_{left}|=\frac{2 c_1}{a_1}-\frac{1} {a_1} \ln  \left( {\frac {a_{1}-a_{2}}{a_{1}+a_
{2}}} \right)$. Then we can define $\int_{-\infty}^{0}|U|^2\mathrm{d}x$ and $\int^{\infty}_{0}|U|^2\mathrm{d}x$ as the particle numbers of two solitons $N_1$ and $N_2$ respectively. They are calculated as
\begin{equation*}
  N_{1}=2(a_1+a_2)+W,\qquad N_{2}=2(a_1+a_2)-W
\end{equation*}
where
\begin{widetext}
\begin{equation*}
  W={\frac { 2(a_1^2-a_2^2)[(a_1+a_2)\sinh( 2 c_{1}-2 c_{2}) + (a_1-a_2)\sinh( 2 c_{1}+2 c_{2})] }{(a_1+a_2)^2\cosh(2c_{1}-2 c_{2})+(a_1-a_2)^2\cosh(2c_{1}+2
c_{2})-4 a_{1}a_{2}\cos(4( {a_{{1}}}^{
2}-{a_{{2}}}^{2})t) }}.
\end{equation*}
\end{widetext}

Based on this, we calculate the exchange particle number between them as $\bigtriangleup N=N_{1}-N_{2}=2 W$. The temporal period for tunneling behavior  is $T_{t}=\frac{\pi}{2(a_1^2-a_2^2)}\approx \frac{2 \pi}{P_1-P_2}$. Especially, it is seen that the tunneling behavior is not a standard cosine form, therefore the tunneling dynamics is different from the linear Josephson oscillation, and it is also a type of nonlinear Josephson oscillation \cite{Wu,Liu}. Then, we can define a critical distance value for the visible tunneling behavior as $d_c$ for which the distance between the two solitons' centers $d=|x_{right}-x_{left}|\geq d_c $, the exchange atoms rate between them satisfies $\frac{\bigtriangleup N_{max}-\bigtriangleup N_{min}}
{N_{1}+N_{2}} \leq 5 \%  $. Then the critical distance value can be calculated numerically with certain values of $a_1$ and $a_2$. It should be noted that the critical distance value depends on the values of $a_1$ and $a_2$, which is related with the profiles of solitons. For example, the critical distance value is calculated as $d_c \approx 6.2$ for the case with $a_1=0.6$ and $a_2=0.5$.

\begin{figure}[htb]
\centering
\includegraphics[height=90mm,width=85mm]{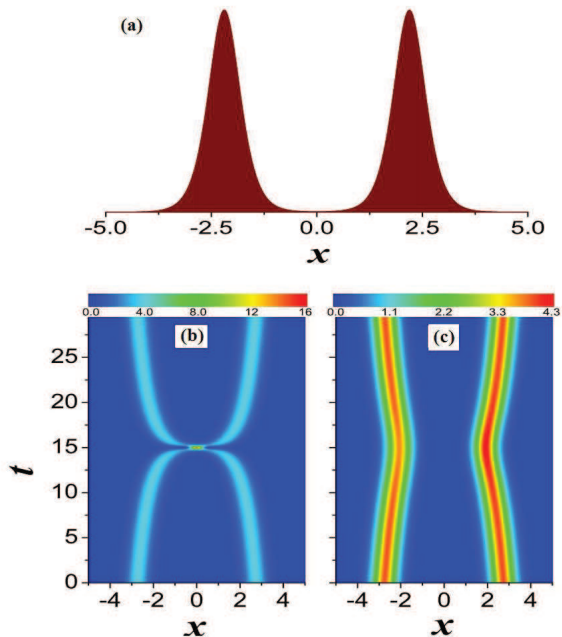}
\caption{(Color online) (a) The initial profiles of two solitons with $a_1=a_2$ and $b_1=b_2$. It is seen that the two solitons have identical profile and energy. Therefore, we call them as resonant interaction.  (b) and (c): The evolution of two solitons with identical energies and identical profiles. (b) $a=b=0$; (c) $a=e^4$, $b=0$. The two solitons demonstrate a resonant interaction which makes them always approach each other one time no matter how far they are located. }\label{fig5}
\end{figure}

\section{A systemic discussion on soliton interaction}
The above studies just demonstrate that tunneling behavior can be visible for the two solitons with different profiles and the identical velocity. Then, what about other cases for two solitons interactions?  We discuss the case for two solitons with nonzero relative velocity firstly, and then study on the case for two solitons with identical profiles and velocities.

\subsection{The case for two solitons with different velocities}
When the two solitons have different initial velocities, they will collide each other. The collision process can be described by the two-soliton solution \cite{Zhao3}. The parameters $a_1$ and $a_2$ determine peak value of solitons respectively, $b_1$ and $b_2$ are related solitons' velocity. $c_1$ and $c_2$ determine the initial locations of solitons. $d_1$ and $d_2$ can be used to vary the relative phase between solitons.
 When the related parameters are chosen, the solution will
present us the dynamics of two solitons directly. The collision of
them can be observed conveniently. Obviously, $a_j$ and $b_j$ determine
soliton's peak and velocity respectively. One can observe interaction between arbitrary two solitons through varying the parameters. There are mainly three different cases for interaction process of solitons with different velocities.

Firstly, when the two solitons have very small relative velocity, namely, $|b_1-b_2|$ has a small value, we can observe the tunneling behavior of two solitons, as shown Fig. 4(a). It is seen that the solitons approach and depart each other very slowly. Then how small the relative velocity is proper for visible tunneling behavior? We can define a critical duration $T_c$, for which solitons' distance is smaller than the critical distance value $d_c$ (for the cases in Fig. 4, the critical distance can be calculated as $d_c\approx 6.2$). $T_c$ can be evaluated by $T_c=\frac{2 d_c} {|v_2-v_1|}$. Low relative velocity can make the duration $\frac{1} {2} T_c$ be longer than the temporal period $ T_t=\frac{2 \pi}{P_1-P_2+\frac{1}{4} (v_2^2-v_1^2)}$ for the tunneling behavior of two solitons with different velocities. We can know that the condition $T_t\leq \frac{1} {2} T_c$ is satisfied for the case in Fig. 4(a) and the tunneling behavior is visible.  Therefore, $T_t\leq\frac{1} {2} T_c$  can be used to clarify relative velocity $v_{c1}$ is needed for the visible tunneling behavior, since the relative velocity determines the time duration $T_c$. From Fig. 4(a), we can also see that the tunneling behavior just can be observed under the condition the distance $d$ between soliton peak locations is smaller than the critical distance $d_c$.

Secondly, the two solitons have larger relative velocity, which brings the duration $T_c$ for two solitons' distance under the critical distance value become shorter. When the condition $T_t\leq \frac{1} {2} T_c$ is not satisfied, the solitons will just demonstrate the particle-like property \cite{Akhmediev2}, there is no visible tunneling behavior, as shown in Fig. 4(b). This is why the tunneling behavior has not been observed for two bright solitons with unequal velocities in the most of previous studies \cite{Zabusky,Segev,Serkin, Serkin1,Zhao,Akhmediev2}. The reasons for why there is also no interference behavior in Fig. 4(b) are discussed in the following paragraph.

Thirdly, when the solitons' relative velocity is increased further, the interference pattern will emerge, as shown in Fig. 4(c). The duration $T_c$ is further decreased and be much shorter than the temporal period $T_t$ of tunneling behavior, which brings that there is no visible tunneling behavior for solitons in Fig. 4(c). The interference pattern property has been analyzed explicitly in \cite{Zhao3}.
The results indicated that the spatial interference period $D=\frac{4 \pi}{v_2-v_1}$  should be smaller than the soliton size $\frac{1} {2} S_{w}$ ($S_w$ is the larger one between two solitons' widths) for visible interference behavior. The condition $D\leq \frac{1} {2}  S_{w}$ can be used to clarify relative velocity $v_{c2}$ is needed for visible interference pattern. This is the reason for that there is no visible interference pattern in Fig. 4(a) and (b). From Fig. 4(c), we can also see that the interference pattern just can be observed under the condition the distance $d$ between soliton peak locations is smaller than the soliton size $s_w$.

\subsection{The case for two solitons with identical profiles and velocities}
When the parameters $a_1=a_2$ and $b_1=b_2$, the two solitons admit  identical profiles and energies (shown Fig. 5 (a)), they will demonstrate a striking different behavior. The soliton's energy can be calculated explicitly by the definition expressions presented in \cite{Martin, T1,T2}. There is no interference pattern, but the tunneling behavior still exist but the tunneling period is infinity which makes the tunneling behavior just happen once, as shown Fig. 5 (b) and (c). The process of two identical solitons with identical energy can be described by the following second-order soliton solution
 $ U(x,t)=\frac { 8\left[  \left( 16 t-b \right) \cosh \left( 2 x \right) +
\mathrm{i} \left( (4 x-a)\sinh \left( 2 x \right)-2
 \cosh \left( 2 x \right)  \right)\right] {{\rm e}^{4 \mathrm{i}t}}}{4\cosh^{2}(2x)+(4x-a)^{2}+(16t-b)^{2}}$,
where $a, b$ are arbitrary real parameters. In Fig.  5 (b) and (c), we can see that the breathing behavior just happen once, and the highest peak depends on the relative phase between two solitons. The difference of relative phase bring the different dynamics in (b) and (c).  It should be noted that the tunneling behavior always happen once even for the distance between two solitons is much larger than the critical distance $d_c$. This comes from the resonance between soliton energies. The velocities of them can  be evaluated by the following approximation analysis. The trajectories of solitons' peak value are along the lines
  $x\pm \frac{1}{4}\ln(2+(4x-a)^2+(16t-b)^2)=0, \text{   as  } x\rightarrow \pm\infty$.
We can see that the velocities of solitons are varied with time, which comes from the resonant nonlinear interaction between solitons. It should  be noted that the velocity evolutions of solitons in this resonant case are different from the case for two solitons with different initial velocities for which the soliton velocity is unchanged when solitons are far apart.
\begin{figure}[htb]
\centering
\includegraphics[height=40mm,width=85mm]{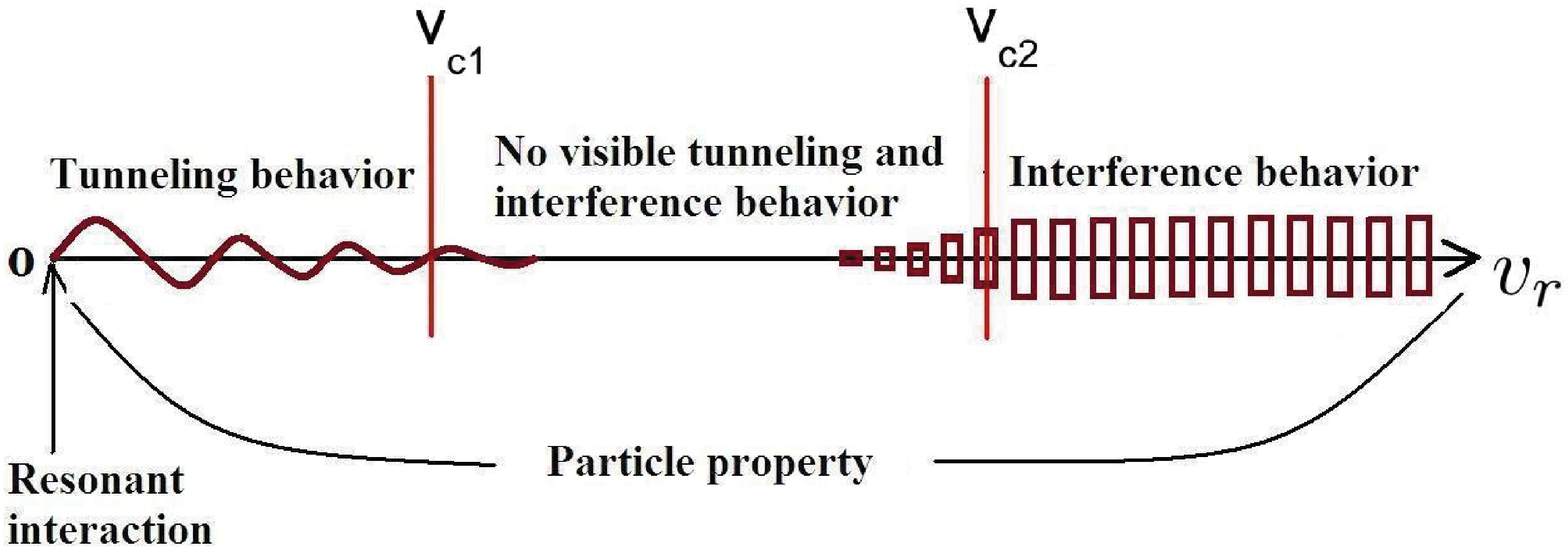}
\caption{(color online) The whole picture for soliton interaction with just considering the effects of relative velocity between solitons. There are mainly four cases: resonant interaction, visible tunneling, no visible tunneling and interference, and visible interference behavior. $v_{c1}$ and $v_{c2}$ can be calculated from the condition \textbf{$T_t =\frac{1} {2} T_c$} and \textbf{$D=\frac{1} {2}  S_{w}$} respectively. } \label{fig7}
\end{figure}

\begin{figure}[htb]
\centering
\includegraphics[height=60mm,width=85mm]{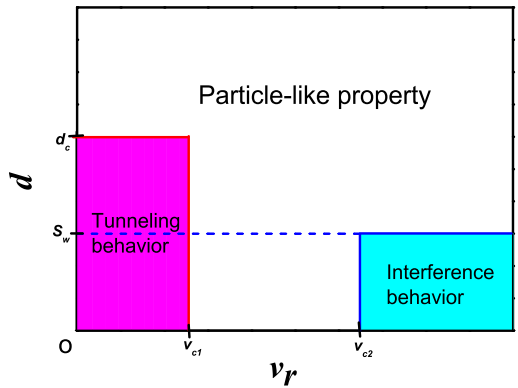}
\caption{(color online) A phase diagram for soliton interaction. The interaction properties of solitons depend on the relative velocity $v_r$ and the distance $d$ between soliton peak locations. There are mainly three cases for solitons with different amplitudes: visible tunneling behavior (the pink regime), particle-like property (the white regime), and visible interference behavior (the mid blue regime). $d_c$ is the critical distance for visible tunneling behavior, which can be evaluated from the soliton amplitude parameters. $S_w$ is the larger one between two solitons' widths. $v_{c1}$  is determined by the condition \textbf{$T_t \leq \frac{1} {2} T_c$}, and $v_{c2}$ is determined by the condition \textbf{$D \leq \frac{1} {2}  S_{w}$}.  Especially, when the two solitons admit identical profile and energy, the critical distance $d_c$ do not stand for this case and they will demonstrate a resonant interaction between solitons. } \label{fig8}
\end{figure}

\subsection{A phase diagram for soliton interaction}
The above discussions show that soliton has both particle and wave properties. The particle-like property can be shown by the elastic collision and always admit certain structure after interaction. The wave property can be shown by tunneling or interference behavior. Tunneling behavior and interference behavior always exist for two solitons interaction, but they can just be visible under some certain conditions. The tunneling behavior is shown clearly under the condition that  \textbf{$T_t \leq \frac{1} {2} T_c$} which determines a critical relative velocity $v_{c1}$ for solitons with certain peak parameters $a_1$ and $a_2$.  The interference behavior is shown clearly under the condition that \textbf{$D\leq \frac{1} {2}  S_{w}$} which determines a critical relative velocity $v_{c2}$ \cite{Zhao3}. When the relative velocity is nonzero, the solitons will always overlap each other for at least one time on the temporal evolution direction. Therefore, the relative velocity are more essential for soliton interactions. We show the cases for soliton interactions in Fig. 6 with just considering the role of relative velocity. When the relative velocity belongs to $[0,v_{c1}]$, the tunneling behavior is visible; when the relative velocity belongs to $(v_{c1},v_{c2})$, both tunneling and interference behavior is invisible; when the relative velocity belongs to $[v_{c2},\infty)$, interference behavior is visible. Especially, when the two solitons admit identical profile and energy, they will approach and depart each other for one time, no matter how far they are located. The critical distance $d_c$ do not stand for this case, we call this as resonant interaction between solitons. From Fig. 2 and Fig. 4, we can see that tunneling behavior or interference pattern just can be visible with the distance between solitons is less than the critical distance $d_c$ or soliton width $S_w$. Therefore, the interaction properties of solitons depend on the relative velocity $v_r$ and the distance $d$ between soliton peak locations, and can be summarized in Fig. 7.   These characters show clearly under which conditions soliton can demonstrate wave property or particle-like property. The results can be used to understand why most of previous studies on soliton interactions do not see the tunneling behavior or interference pattern. The tunneling period and rate or interference pattern periods are calculated analytically. This will further deepen our realization and understanding of bright soliton greatly.

\section{Conclusion and Discussion}

We demonstrate that the breathing behavior of bound state solitons comes from the tunneling dynamics of matter wave in an effective double-well potential induced by the two solitons. The tunneling rate and oscillation period are derived analytically and exactly. The results provide new possibilities to observe nonlinear Josephson oscillation of cold atoms based on two bright matter wave solitons with the zero initial relative velocity. This can be seen as  the mechanism for breathing bound state of two solitons, in contrast to the modulational instability mechanism for Akhmediev breather and K-M breather.

Furthermore, we present a phase diagram for two solitons interaction which admits tunneling property, particle-like property,  interference property and a resonant interaction case. The explicit conditions for them are clarified based on the critical distance $d_c$ and spatial interference period $D$. The results here can be extended to three or more solitons cases and vector soliton system \cite{z2,z3,Becker,Bludov}.

 Recent studies have shown that the interference pattern holds great promise
for precision measurements \cite{Polo,22,Martin}, including measurements
of gravity \cite{23,26}, rotations and magnetic field
gradients \cite{29}, and other quantum superpositions \cite{30,31}. The tunneling behavior here could be used to test quantum coherent degree between solitons and provide some implications on quantum entanglement state preparations.

\begin{acknowledgements}
This work is supported by National Natural Science Foundation of
China (Contact No. 11405129, 11404259).
\end{acknowledgements}

\end{document}